\documentclass[reprint,aps,ams,amsmath,prx,twocolumn,longbibliography]{revtex4-1}
\usepackage{graphics}
\usepackage{graphicx}
\usepackage{epsfig}
\usepackage{amsmath}
\usepackage{amssymb}
\usepackage{amsfonts}
\usepackage{bm}
\usepackage{color}
\usepackage{bbm}
\usepackage{hyperref}
\usepackage[normalem]{ulem}
\usepackage{comment}
\usepackage{soul}
\usepackage[OT1]{fontenc}

\begin{document}

\title{Spin mechanism of drag resistance in strongly correlated electron liquids}

\author{Dmitry Zverevich}
\affiliation{Department of Physics, University of Wisconsin-Madison, Madison, Wisconsin 53706, USA}

\author{Ilya Esterlis}
\affiliation{Department of Physics, University of Wisconsin-Madison, Madison, Wisconsin 53706, USA}

\author{Alex Levchenko}
\affiliation{Department of Physics, University of Wisconsin-Madison, Madison, Wisconsin 53706, USA}

\date{August 25, 2024}

\begin{abstract}
We investigate the effect of Coulomb drag resistance in a bilayer system of strongly-correlated electron liquids magnetized by an in-plane field employing the framework of hydrodynamic theory. We identify a mechanism for drag magnetoresistance, which physically arises from the spin diffusion driven by fluctuations of the spin currents within a partially spin-polarized fluid. This effect is further enhanced by acoustic and optic plasmon resonances within the bilayer, where hydrodynamic plasmons are driven by fluctuating viscous stresses. We express the drag magnetoresistivity in terms of the intrinsic dissipative coefficients and basic thermodynamic properties of the electron fluid. Our results are derived nonperturbatively in interaction strength and do not rely on assuming Fermi-liquid behavior of the electron liquid, and applicable also in the regimes 
of semiquantum and highly correlated classical fluids.
\end{abstract}

\maketitle

\section{Introduction and motivation} 

In two-dimensional electron systems (2DES), the effects of Coulomb interaction on thermodynamic and transport properties become especially pronounced in the limit of dilute particle density \cite{Ando:RMP82,Gao:RMP10}. The strength of these interactions can be quantified by the dimensionless parameter $r_s=V_{\text{C}}/E_{\text{F}}$, which is defined as the ratio of the Coulomb energy $V_{\text{C}}$ to the Fermi energy $E_{\text{F}}$. As a function of the particle density $n$, $r_s$ can be expressed as $r_s=1/\sqrt{\pi na^2_{\text{B}}}$, where $a_{\text{B}}=\epsilon/me^2$ is the Bohr radius in the material with dielectric constant $\epsilon$ and effective mass $m$ (hereafter, Planck's constant $\hbar=1$). For values of $r_s<1$, the system forms a weakly interacting Fermi gas, which is well described in the framework of Fermi-liquid (FL) theory. 

The regime of strong electron correlations extends over the broad range of $1<r_s<r_{sc}$ up to the critical value $r_{sc}$, marking the Wigner crystallization of the electron system \cite{Wigner}. State-of-the-art numerical methods, such as quantum Monte Carlo methods \cite{Tanatar:PRB89,Drummond:PRL09}, provide reliable estimates that the quantum critical point (QCP) associated with the quantum liquid-to-Wigner solid transition occurs at $r_{sc}\simeq 35$ consistent with the experimental observations, see e.g. Ref. \cite{Yoon:PRL99}. At finite temperature, the Wigner crystal (WC) melts. In the $T$-versus-$r_s$ phase diagram, various mesoscopic and microemulsion phases of electron liquid were predicted, including different forms of magnetism of a WC \cite{Spivak:PRB04,Drummond:PRL09,Ceperley:PRL09}. However, despite decades of efforts, we know little beyond the basic thermodynamic properties of these electronic states.  
 
Transport measurements of Coulomb drag provide a useful probe of correlation effects in electron bilayers (see review \cite{Narozhny:RMP16} and references therein). For example, drag resistance $\rho_{\text{D}}$ reveals signatures of superconducting fluctuations \cite{Giordano:PRB94}, interlayer exciton condensation \cite{Lilly:PRL09,Lilly:PRB09}, and exquisite properties of quantum Hall liquids \cite{Kellogg:PRL02,Shayegan:PRB09}. In the present paper, we focus on the physics of nonlocal transport in 2DES at large $r_s$, motivated in part by a long-standing problem: the unusual temperature ($T$) and magnetic field ($H$) dependences of drag resistance $\rho_{\text{D}}(T,H)$ observed in bilayer $p$-type GaAs heterojunction devices \cite{Pillarisetty:PRL02,Pillarisetty:PRL03, Pillarisetty:PRL05}. 
For the problem in question, the field dependence concerns the spin effect rather than the orbital effect. We describe a particular mechanism of spin-induced magnetodrag resistance, derive the corresponding results that are nonperturbative in interaction, and, in principle, applicable to non-Fermi liquids.

This paper is structured as follows. In Sec. \ref{sec:Background}, we begin with a brief summary of the experimental situation and formulate the problem. We also review the relevant existing theories devoted to this particular topic in parallel. This enables us to place our paper in the context of existing studies. In Sec. \ref{sec:Hydro}, we formulate our approach to the problem of drag in 2DES with $r_s\gg1$. It is based on the theory of hydrodynamic fluctuations. To describe magnetotransport properties, we generalize this theory to include fluctuations of the spin density and spin currents in the presence of partial spin polarization induced by the external field. In Sec. \ref{sec:Drag}, we apply this theory to the calculation of drag magnetoresistance. Finally, in Sec. \ref{sec:Summary}, we provide summary of main findings and concluding remarks.

%#################################################################################################################
%#################################################################################################################
%#################################################################################################################
%#################################################################################################################
%#################################################################################################################

\section{Theoretical and experimental background}\label{sec:Background} 

The original measurements of drag resistance in weakly correlated electron bilayers \cite{Gramila:PRL91}, with $r_s\lesssim1$, are in both qualitative and quantitative agreement with existing Fermi-liquid theories of drag \cite{Jauho:PRB93,Zheng:PRB93,Kamenev:PRB95,Flensberg:PRB95}. The drag resistance is given by (up to a numerical factor)  
\begin{equation}\label{eq:rho-D-FL}
\rho_{\text{D}}\simeq \frac{h}{e^2}\frac{1}{(k_{\text{F}}d)^2(k_{\text{TF}}d)^2}\left(\frac{T}{E_\text{F}}\right)^2.
\end{equation}
Here, $d$ is the interlayer separation, $k_{\text{F}}$ is the Fermi momentum, and $k_{\text{TF}}=2\pi e^2\nu/\epsilon\sim r_s k_{\text{F}}$ is the inverse Thomas-Fermi screening radius, with $\nu$ being the single-particle density of states. For simplicity, we assume identical layers with matched densities and we will focus only on the expressions in the clean limit $k_{\text{F}}l\gg1$, where $l$ is the elastic mean-free path. 
This limit is particularly relevant to bilayer devices at large $r_s$ since drag experiments in these systems were performed on high-mobility samples whose conductances are significantly larger than conductance quantum $\sim e^2/h$. Therefore, disorder renormalizations of scattering and screening effects can be neglected. 

\subsection{Temperature dependence} 

The quadratic temperature dependence of drag resistance in Eq. \eqref{eq:rho-D-FL} can be understood from a simple phase-space argument: for two-particle collisions, there are $\propto T$ states near the Fermi level per layer that are susceptible to scattering when $T\ll E_{\text{F}}$. The scaling of the drag resistance with the interlayer separation, $\propto 1/d^4$, also follows naturally from the properties of the dynamically screened Coulomb interaction. In this limit, drag is dominated by excitations in the particle-hole continuum, and for a long time, it was believed that Eq. \eqref{eq:rho-D-FL} applies at all temperatures below the Fermi energy. For this reason, deviations from the $\propto T^2$ behavior are typically attributed to signatures of non-Fermi liquid physics. Indeed, such arguments were put forward since subsequent measurements in strongly correlated bilayers with $r_s\gg1$ not only found a non-quadratic temperature dependence of $\rho_{\text{D}}$ over a broad range of temperatures, but also a drag resistance magnitude that is two to three orders of magnitude larger than expected based on Fermi liquid theory \cite{Pillarisetty:PRL02}. 

An attempt to resolve these discrepancies was put forward in Ref. \cite{Stern:PRL03}. The theory was based on a standard perturbative expression for the drag resistivity valid at $r_s\sim 1$, using an improved Hubbard approximation for the single-layer polarization operator and the experimentally measured density dependence of the single-layer conductivity. However, it should be noted that attempting to extrapolate usual expressions to the limit of large $r_s\gg1$ is uncontrolled and, strictly speaking, lacks solid justification. This approach could also lead to unphysical predictions regarding the screening properties of the electron liquid.

An alternative explanation for the strong enhancement of drag resistance at large $r_s$ was proposed in Ref. \cite{Spivak:PRB05}, based on the picture of microemulsion phases. At the level of mean-field theory, these phases can be characterized as a mixture of microphase-separated regions of liquid and crystal. Practical considerations for the drag resistance were given for the bubble phase, which can be viewed as a suspension of WC islands floating in a uniform Fermi liquid. Under such assumptions, an increased drag can be attributed to the Pomeranchuk effect: because the WC has a higher spin entropy density than a Fermi liquid, the areal fraction of the WC increases with temperature. This fraction can be determined from thermodynamic considerations by examining the free energy difference between the uniform WC and FL phases, which scales linearly with temperature as $\propto nT\ln 2$. This explanation is plausible, although its underlying theory is purely phenomenological.
For completeness, we should also mention Ref. \cite{Braude:PRB01}, where an asymmetric WC-FL bilayer was considered with the WC phase pinned by disorder. In this case, drag resistance was shown to scale as $\rho_{\text{D}}\propto T^4$. This result is analogous to the Bloch law for the contribution of electron-phonon interaction to the resistivity and drag \cite{Bonsager:PRB98}. However, this specific mechanism leads to a significant decrease in transresistivity and requires disorder, thus unlikely to be relevant in application to experiments in high-mobility systems.

It should be emphasized, that even in a weakly-correlated regime, $r_s\sim1$, where the framework of FL applies, the temperature dependence of drag resistance is not simply quadratic. It turns out that Eq. \eqref{eq:rho-D-FL} is valid only at lowest temperatures when $T<E_{\text{F}}/(k_{\text{F}}d)$. Note that for GaAs quantum wells typically $k_{\text{F}}d>1$. At slightly higher temperatures, $E_{\text{F}}/(k_{\text{F}}d)<T<E_{\text{F}}/\sqrt{k_{\text{F}}d}$, drag resistance is given by a different expression such that $\rho_{\text{D}}\propto T$.   
It was implicit in the analysis of Ref. \cite{Jauho:PRB93}, but not clearly emphasized, and later rediscovered in Ref. \cite{Chen:PRB15}. As compared to Eq. \eqref{eq:rho-D-FL}, in this limit, the kinematics of collisions near the threshold of particle-hole continuum for typical frequencies and wave numbers where $\omega\sim v_{\text{F}}q$ changes the drag resistance to be linear in temperature. The upper limit in the applicability condition of the $\rho_{\text{D}}\propto T$ behavior, namely $\sim E_{\text{F}}/\sqrt{k_{\text{F}}d}$, marks the energy scale at which the intralayer mean-free path because of electron collisions $\ell(T)$ becoming comparable or smaller than the interlayer distance $d$. This regime defines the onset of collision-dominated transport. At even higher temperatures, $E_{\text{F}}/\sqrt{k_{\text{F}}d}<T<E_{\text{F}}/\sqrt[4]{k_{\text{F}}d}$, drag resistance is dominated by plasmons and $\rho_{\text{D}}\propto T^3$. The energy scale $\sim E_{\text{F}}/\sqrt[4]{k_{\text{F}}d}$ corresponds to the hydrodynamic regime of plasmons, when their frequency $\omega_{\text{pl}}$ at the typical wave-numbers, $q\sim1/d$, becomes comparable or smaller than the rate of electron collisions $\tau^{-1}$. Above this temperature, but still below $E_{\text{F}}$, one finds drag decreasing with temperature as $\rho_{D}\propto 1/T$ \cite{Apostolov:PRB14}.       

\subsection{Magnetic field dependence} 

The observed dependence of Coulomb drag at large $r_s$ on spin polarization induced by the application of an in-plane magnetic field, as described in Refs. \cite{Pillarisetty:PRL03, Pillarisetty:PRL05}, raises additional questions and challenges for the theory. Indeed, with an increase in  $H$, the temperature dependence is significantly suppressed: whereas $\rho_D\propto T^{\alpha_T}$ with $\alpha_T\simeq 3$ in the zero-field limit, $\alpha_T\simeq 1$ for $H\sim H^*$, where $H^*$ corresponds to the field required to fully polarize the electron spins. The magnitude of drag magnetoresistance increases rapidly with $H$, by as much as an order of magnitude, and then tends to saturate for  $H>H^*$. The fit to the data at low fields suggests a quadratic field dependence, $\rho_{\text{D}}\propto H^2$. The change in the temperature dependence induced by the finite field is monitored through the $\rho_{\text{D}}/T^2$ ratio, which changes from increasing with $T$ to decreasing with $T$ very close to $H^*$. 
Perhaps an even more striking feature is that both the temperature and magnetic field dependences of single-layer magnetoresistance 
$\rho$ and drag magnetoresistance $\rho_D$ appear qualitatively similar. This is unexpected \textit{a priori} since these kinetic coefficients describe completely different transport properties. This similarity suggests a unifying mechanism for the spin-polarization dependence of both $\rho$ and $\rho_{\text{D}}$.

The theory presented in Ref. \cite{Hwang:PRB05} addressed the experimental observations of $\rho_{\text{D}}(T,H)$ based on the random-phase approximation (RPA) with an extrapolation of computed results to large values of $r_s$. However, this approach is not physically justified as it extends the RPA theory beyond the limit of its validity. In contrast, in the theory of microemulsion phases, a strong increase in drag with the field can be qualitatively explained by the argument that spins are substantially more polarizable in the Wigner crystal (WC) phase \cite{Spivak:PRB05}. It should be noted that there is strong experimental support for the enhancement of spin susceptibility in dilute 2DES \cite{Klapwijk:PRL01,Zhu:PRL03,Tutuc:PRB03}.

\subsection{Density dependence} 

The observed density dependence of the drag resistance is equally intricate as discussed in detail in Refs. \cite{Pillarisetty:PRL03, Pillarisetty:PRL05}. To put this in perspective, recall that the FL theory predicts $\rho_{\text{D}}\propto 1/n^3$, per Eq. \eqref{eq:rho-D-FL}. However, the experiment reveals a much stronger dependence, $\rho_{\text{D}}\propto 1/n^5$, which was deduced from the data collapse in bilayers with $r_s\gg1$ and matched density. This power-law, found at low temperatures, is mostly unaffected by a small parallel magnetic field. In general, experiments found that at zero field and for $T\lesssim E_{\text{F}}/2$, $\rho_{\text{D}}$ follows approximately a $\propto 1/n^{5/2}_{i}$ dependence upon either layer density $n_{1,2}$. For mismatched densities at zero field, drag resistance is a completely monotonic decaying function of the density ratio, when density is swept in one layer by a gate voltage, while the other is kept fixed. In contrast, at higher fields, a different behavior is observed. A clear enhancement of $\rho_{\text{D}}$ from $H$ shows a nonmonotonic behavior as a function of density ratio between the layers, exhibiting a local maximum at essentially matched densities.        

\subsection{Hydrodynamic description} 
  
In this paper, we develop a theory of drag in bilayers of spin-polarized electron liquids. Our approach is based on the hydrodynamic description of transport in the regime of strong electron correlations originally described in Ref. \cite{AKS}. This approach can be justified in high-mobility conductors within a range of temperatures and sample purities where the electron-electron mean-free path is small compared to the length scales over which momentum and energy conservation of electron systems are violated. An extension of the theory in Ref. \cite{AKS} to include hydrodynamic fluctuations and compute drag resistance in zero field was given in Refs. \cite{Apostolov:PRB14,Chen:PRB15,Holder:PRB19}. The orbital effect of the magnetic field was considered in Refs. \cite{Patel:PRB17,Apostolov:PRB19}. Here, we focus on the spin effect and incorporate fluctuations of the spin density and spin current into the main set of hydrodynamic equations, supplemented by the fluctuation-dissipation relations. This enables us to consider both drag and single-layer resistivities from a unified perspective. Our results are as general as any hydrodynamic theory and rely only on the applicability conditions of the hydrodynamic approximation. We analyze our results for the temperature, field, and density dependence of drag resistance in light of experimental findings \cite{Pillarisetty:PRL02,Pillarisetty:PRL03, Pillarisetty:PRL05}.

In closing this section, for clarity, we comment on one terminological detail. We use the term spin drag to describe Coulomb drag resistance in a bilayer system. In the literature, there also exists a term spin Coulomb drag, which describes a completely different phenomenon \cite{DAmico:PRB00,DAmico:PRB03}. Spin Coulomb drag refers to the mutual friction between spin-polarized electron populations in a single layer, which can be observed in spin-valve devices and 2D electron systems (2DES) \cite{Weber:2005}. The theory of spin Coulomb drag relies on the key assumption that the spin-up and spin-down components of the electron system have distinct drift velocities. In the hydrodynamic regime dominated by electron collisions that we are interested in, this is not possible as both electron subsystems develop a single hydrodynamic velocity.

%#################################################################################################################
%#################################################################################################################
%#################################################################################################################
%#################################################################################################################
%#################################################################################################################

\section{Hydrodynamic theory of Coulomb drag for spin-polarized 2DES}\label{sec:Hydro} 

\subsection{Main equations}

A hydrodynamic description of electron transport is based on the existence of slow variables associated with conserved quantities \cite{AKS}. The conservation of the total particle number is captured by the continuity equation  
\begin{equation}\label{eq:dtn}
\partial_t\delta n+\bm{\nabla}\cdot \delta\bm{j}_n=0. 
\end{equation}
Here we present it in the linearized form written for the fluctuation components of the particle density $\delta n(\bm{r},t) $ and corresponding particle current density 
\begin{equation}
\delta\bm{j}_n=\bm{v}\delta n+n \delta\bm{v},
\end{equation} 
where $\bm{v}$ and $\delta\bm{v}(\bm{r},t)$ is the hydrodynamic velocity of the flow and its fluctuating component respectively. In equilibrium $\bm{v}=0$ and particle density is uniform. 

The energy conservation is usually replaced by an equivalent evolution equation for the entropy density \cite{LL-V6} 
\begin{equation}\label{eq:dts}
nT(\partial_t+\bm{v}\cdot\bm{\nabla})\delta s+\bm{\nabla}\cdot\delta\bm{j}_q=0. 
\end{equation}
where $T$ is the equilibrium temperature. The fluctuation of the heat current $\delta\bm{j}_q$ is related to the temperature fluctuation $\delta T(\bm{r},t)$ by Fick's law 
\begin{equation}
\delta\bm{j}_q=-\kappa\bm{\nabla}\delta T+\delta\bm{g},
\end{equation} 
where $\kappa$ is the thermal conductivity of the fluid. 
The second term in the current fluctuation $\delta\bm{g}(\bm{r},t)$ is the stochastic Langevin heat flux whose correlation function is given by the fluctuation-dissipation-relation \cite{LL-V9}
\begin{equation}
\langle g_i(\bm{r},t)g_k(\bm{r}',t')\rangle=2\kappa T^2\delta_{ik}\delta_{\bm{r}-\bm{r}'}\delta_{t-t'},
\end{equation}
where $i,k$ are Cartesian indices for the coordinates in 2D plane and $\langle\ldots\rangle$ denotes averaging over the thermal fluctuations.  

We add now a conservation law for the spin current \cite{Forster}. To this end, we assume that the spin-orbit interaction is weak, so that the spin component $\sigma$ along the magnetic field applied parallel to the plane of 2DES is approximately conserved. This is reasonable assumption since spin relaxation in GaAs is sufficiently long. We thus have for the spin fluctuation
\begin{equation}\label{eq:dtsigma}
(\partial_t+\bm{v}\cdot\bm{\nabla})\delta\sigma+\bm{\nabla}\cdot\delta\bm{j}_\sigma=0. 
\end{equation}
It is important to stress that only the spin component along the magnetic field appears in this equation since the other spin components are not conserved because of spin precession. The fluctuation of the spin current density 
\begin{equation}
\delta\bm{j}_\sigma=-\sigma_s\bm{\nabla}\delta\mu_\sigma+\delta\bm{\varsigma}
\end{equation}
is related to the gradient of the spin chemical potential fluctuation $\delta\mu_\sigma$ via the spin conductivity $\sigma_s$. The latter can be expressed in terms of the spin diffusion constant $D_\sigma$ via the Einstein relation $\sigma_s=\chi D_\sigma$, where $\chi$ is the magnetic susceptibility. In the expression for the spin current, we neglected the term corresponding to the spin thermocurrent that must be present by the Onsager principle. Its contribution is important for spin-caloric and spin-drag thermal resistances near charge neutrality \cite{AL:Spin-Drag,AL:Spin-Calorics}, but is expected to be small for electrical properties at high doping as compared to the spin diffusion term. In accordance with the fluctuation-dissipation theorem, the random flux of Langevin spin currents correlates to the spin conductivity 
\begin{equation}\label{eq:Langevin-spin}
\langle \varsigma_i(\bm{r},t)\varsigma_k(\bm{r}',t')\rangle=2\sigma_s T\delta_{ik}\delta_{\bm{r}-\bm{r}'}\delta_{t-t'},
\end{equation}

The continuity equation for momentum density has the standard form 
\begin{equation}\label{eq:dtp}
mn(\partial_t+\bm{v}\cdot\bm{\nabla})\delta\bm{v}+\bm{\nabla}\cdot\delta\hat{\mathbf{\Pi}}=0. 
\end{equation}
Here fluctuations of the momentum flux tensor 
\begin{equation}\label{eq:deltaPi}
\delta\hat{\Pi}_{ik}=(\delta P+en\delta\Phi)\delta_{ik}-\delta\hat{\Sigma}_{ik}
\end{equation}
comprise of local hydrodynamics fluctuations in the pressure of a fluid $\delta P$, fluctuations of the long-ranged Coulomb potential $\delta\Phi$, and fluctuations of viscous stresses
\begin{equation}
\delta\hat{\Sigma}_{ik}=\eta(\partial_k\delta v_i+\partial_i\delta v_k)+(\zeta-\eta)\delta_{ik}\bm{\nabla}\cdot\delta\bm{v}+\delta\hat{\Xi}_{ik}.
\end{equation}
The latter are  expressed in terms of the spatial gradients $\partial_i=\partial/\partial x_i$ of the velocity field components $\delta v_k$, where $\eta$ and $\zeta$ are, respectively, the shear and bulk viscosities. The random fluxes of fluctuating viscous stresses in the fluid are described by the correlation function of the form
\begin{align}\label{eq:Langevin-visc}
&\langle\delta\hat{\Xi}_{ik}(\bm{r},t)\delta\hat{\Xi}_{lm}(\bm{r}',t')\rangle=\nonumber \\
&2T[\eta(\delta_{il}\delta_{km}+\delta_{im}\delta_{kl})+(\zeta-\eta)\delta_{ik}\delta_{lm}]\delta_{\bm{r}-\bm{r}'}\delta_{t-t'}.
\end{align}

The hydrodynamic description applies to any liquid type whether it is Fermi liquid or not. Microscopic properties of the liquid manifest themselves via the temperature and density dependence of the intrinsic dissipative coefficients as well as thermodynamic quantities, and the equation of state. Noteworthy, in high-mobility and low-density 2DES the momentum and energy-relaxing electron-phonon scattering is still weak even at $T> E_{\text{F}}$ \cite{Gao:PRL05}. Therefore, the hydrodynamic description applies from very short distances of the order of interelectron spacing for both semiquantum liquids at $E_{\text{F}}<T<\sqrt{r_s}E_{\text{F}}$, or highly correlated classical fluid at $\sqrt{r_s}E_{\text{F}}<T<r_sE_{\text{F}}$.  It should be mentioned that there is no established theory in these regimes, but a conjecture concerning the $T$ dependences of $s, \kappa, \eta, \zeta$ was put forward in Ref. \cite{Kivelson:AP06}. We have also our own results for some of these quantities that will be reported in a separate publication. 

%%%%%%%%%%%%%%%%%%%%%%%%%%%%%%
%%%%%%%%%%%%%%%%%%%%%%%%%%%%%%
\begin{figure}[t!]
\includegraphics[width=0.9\linewidth]{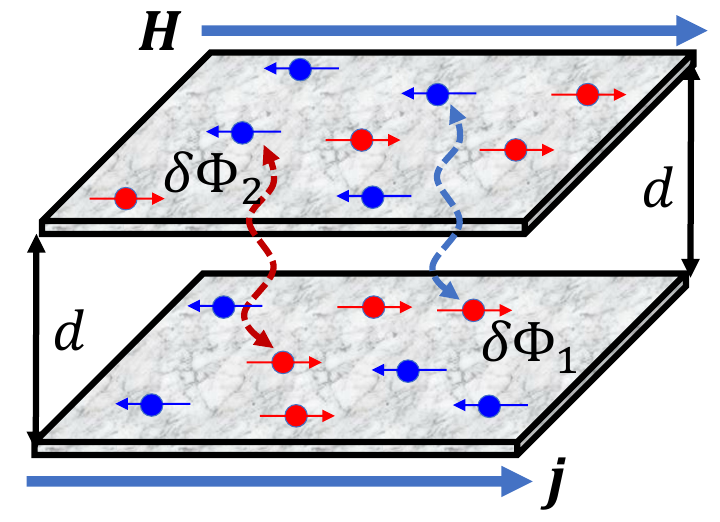}
\caption{Schematic representation of the Coulomb drag setup consisting of two interactively coupled 2DES separated by a distance $d$. 
Both layers are magnetized by an in-plane magnetic field $H$. A steady hydrodynamic flow is imposed in the active layer with the current density $\bm{j}$. Coulomb interaction between the particles is depicted by wavy lines with the corresponding dynamically fluctuating potentials marked as $\delta\Phi_{1,2}$ per Eq. \eqref{eq:delta-Phi}. The electrons (holes) are depicted as circles with arrows showing spin polarization. The underlying background of each layer represents a snapshot of inhomogeneous thermally driven particle and spin density fluctuation profiles.}
\label{fig:bilayer}
\end{figure}
%%%%%%%%%%%%%%%%%%%%%%%%%%%%%%
%%%%%%%%%%%%%%%%%%%%%%%%%%%%%%

\subsection{Solution for a bilayer setup}

Consider now a bilayer geometry relevant for the Coulomb drag setup Fig. \ref{fig:bilayer}. For simplicity we assume identical 2DES. To describe this system we need to duplicate the number of equations describing fluctuations of particle density Eq. \eqref{eq:dtn}, entropy density Eq. \eqref{eq:dts}, spin density Eq. \eqref{eq:dtsigma}, and hydrodynamic velocity Eq. \eqref{eq:dtp} in each of the layer and their coupling. For hydrodynamic model to be applicable, we are constrained to work under the tacit condition that interlayer separation exceeds the intralayer mean-free path because of electron collisions $d\gg\ell$. Since $\ell$ diverges in the low temperature limit, hydrodynamics sets in at the intermediate range of temperatures. For a Fermi liquid, the onset of collision-dominated regime is marked by the temperature $T\gtrsim E_{\text{F}}/\sqrt{k_{\text{F}}d}$. In the following, we denote density fluctuations in each layer as $\delta n_{1,2}(\bm{r},t)$, and similarly for all other quantities. 

To solve the equations we use Fourier transform $\{\delta n, \delta s, \delta\bm{v}, \delta\sigma\}\propto\exp(-i\omega t+i\bm{qr})$. From Eq. \eqref{eq:dtn} we get in the active layer  
\begin{equation}\label{eq:FT-dtn}
[\omega-(\bm{q}\cdot\bm{v})]\delta n_1=n(\bm{q}\cdot\delta\bm{v}_1).
\end{equation}  
In the passive layer the equation is the same except $\bm{v}=0$. In other words, we assume open circuit boundary condition, where the drag force exerted from the active layer onto the passive layer is compensated by a force from the built-in electric field in the passive layer in a stationary fluid. Analogously, from Eq. \eqref{eq:dts} we obtain 
\begin{equation}\label{eq:FT-dts}
nT[-i\omega+i(\bm{q}\cdot\bm{v})]\delta s_1=-\kappa q^2\delta T_1-i(\bm{q}\cdot\delta\bm{g}_1), 
\end{equation}
while from Eq. \eqref{eq:dtsigma} we similarly find 
\begin{equation}\label{eq:FT-dtsigma}
[-i\omega+i(\bm{q}\cdot\bm{v})]\delta\sigma_1=-\sigma_s q^2\delta\mu_{\sigma1}-i(\bm{q}\cdot\delta\bm{\varsigma}_1). 
\end{equation}
Finally, from the momentum balance equation \eqref{eq:dtp}
\begin{equation}\label{eq:FT-dtp}
[-i\omega+i(\bm{q}\cdot\bm{v})]\delta\bm{v}_1=-\frac{i\bm{q}}{m}e\delta\Phi_1-\frac{i\bm{q}}{mn}\delta P_1+\frac{i(\bm{q}\cdot\delta\hat{\bm{\Sigma}}_1)}{mn}.
\end{equation}

The main mechanism of coupling between the layers is by Coulomb potential that includes both fluctuations of the density in the layer-1 as well as dynamically screened fluctuations of the density in the layer-2. In layer-1 it takes the form 
\begin{equation}\label{eq:delta-Phi}
\delta\Phi_1(\bm{q},\omega)=\frac{2\pi e}{\epsilon q}\left[\delta n_1(\bm{q},\omega)+e^{-qd}\delta n_2(\bm{q},\omega)\right],
\end{equation}
and similarly in the other layer $\delta\Phi_2$, which is obtained by simply interchanging the indices. The fluctuations of pressure in the local part of the momentum flux tensor Eq. \eqref{eq:deltaPi} has several contributions
\begin{equation}\label{eq:deltaP}
\delta P_1=(\partial_nP)\delta n_1+(\partial_sP)\delta s_1+(\partial_\sigma P)\delta\sigma_1.
\end{equation} 
The first term can be neglected as compared to the corresponding term in the fluctuations of the Coulomb potential that dominates in the long wave length limit, $q\to0$, so that $en\delta\Phi\gg(\partial_nP)\delta n$. The second term leads to the interlayer density coupling via the thermal expansion of the fluid mediated by thermal fluxes. Calculation shows that the contribution of this term to drag resistivity is subleading and, therefore, can be neglected. In this approximation Eq. \eqref{eq:FT-dts} decouples and plays no role in the remaining analysis. The last term in Eq. \eqref{eq:deltaP} must be retained as it captures the main effect of spin polarization on the equation of state of the fluid.    

The steady current $\propto n\bm{v}$ in the active layer exerts the drag on the passive layer. Using the Poisson equation to relate the potential to density fluctuations, and ignoring the intralayer thermal forces, we can express the drag force, 
\begin{equation}\label{eq:F-D}
\bm{\mathcal{F}}_{\text{D}}=\int (-i\bm{q})\left(\frac{2\pi e^2}{\epsilon q}\right)e^{-qd}\mathcal{D}(\bm{q},\omega)d\Gamma_{q\omega},
\end{equation}
in terms of the interlayer density-density correlation function
\begin{equation}\label{eq:D}
\mathcal{D}(\bm{q},\omega)=\left\langle\delta n_1(\bm{q},\omega)\delta n_2(-\bm{q},-\omega)\right\rangle,
\end{equation}
where the integration expands over the phase space $d\Gamma_{q\omega}=d\omega d^2q/(2\pi)^3$. In the linear response, drag resistance is the proportionality coefficient between the electric field in the passive layer and the current in the active layer $\bm{E}_2=\rho_{\text{D}}\bm{j}_1$. From the force balance condition on an element of the fluid we have $\bm{\mathcal{F}}_{\text{D}}=en\bm{E}_2$, while $\bm{j}_1=en\bm{v}$, therefore, 
knowing the drag force one readily finds the drag resistivity
\begin{equation}\label{eq:rho-D-def}
\rho_{\text{D}}=\frac{\bm{\mathcal{F}}_{\text{D}}}{e^2n^2\bm{v}}. 
\end{equation}    

The remaining technical task is to determine the linear in $\bm v$ part of the drag force, or equivalently the density structure factor in Eq. \eqref{eq:D}. To prepare for this computation we proceed as follows. (i) We multiply Eq. \eqref{eq:FT-dtp} by an extra power of $\bm{q}$ and express the product $(\bm{q}\cdot\delta\bm{v}_1)$ in terms of $\delta n_1$ with the help of the continuity equation \eqref{eq:FT-dtn}. In this step, we eliminate fluctuations of hydrodynamic velocity from the system of equations. We thus left with the coupled fluctuations for $\delta n$ and $\delta\sigma$. (ii) In Eq. \eqref{eq:FT-dtsigma} we write $\delta\mu_\sigma=(\partial_\sigma\mu_\sigma)\delta\sigma+(\partial_n\mu_\sigma)\delta n$, solve for $\delta\sigma_{1,2}$ in terms of $\delta n_{1,2}$, and insert that expression into Eq. \eqref{eq:FT-dtp}. This leaves us with two coupled algebraic equations for $\delta n_{1,2}$. (iii) We make a linear transformation to the basis of symmetrized/antisymmetrized density modes $\delta n_\pm=\delta n_1\pm\delta n_2$. In this representation our algebraic equations decouple, and we solves them perturbatively to the linear order in $\bm{v}$. This gives a result 
\begin{subequations}\label{eq:delta-n-sol}
\begin{equation}
\delta n_\pm(\bm{q},\omega)=\delta n^{(0)}_\pm(\bm{q},\omega)+\delta n^{(1)}_\pm(\bm{q},\omega),
\end{equation}
\begin{equation}
\delta n^{(0)}_\pm=\frac{\omega+i\omega_{\sigma}}{m\Pi_\pm}\bm{q}\cdot(\bm{q}\cdot\delta\hat{\bm{\Xi}}_\pm)-\frac{q^2\partial_\sigma P}{m\Pi_\pm}(\bm{q}\cdot\delta \bm{\varsigma}_\pm)
\end{equation}
\begin{equation}
\delta n^{(1)}_\pm=\frac{(\bm{q}\cdot\bm{v})}{2\Pi_\pm}\sum_{i=\pm}\left[\Gamma_i\delta n^{(0)}_i-\frac{\bm{q}\cdot(\bm{q}\cdot\delta\hat{\bm{\Xi}}_i)}{m}\right]
\end{equation}
\end{subequations}    
In this solution, $\delta n^{(0)}_\pm$ describe the equilibrium density fluctuations driven by thermal fluctuations of viscous stresses $\delta\hat{\bm{\Xi}}$ and spin currents $\delta\bm{\varsigma}_\pm$. The additional terms, $\delta n^{(1)}_\pm$, describe nonequilibrium parts of density fluctuations advected by the hydrodynamic flow. In the above expressions, we introduced the following notations 
\begin{subequations}\label{eq:Pi-Gamma}
\begin{equation}
\Pi_\pm=-(\omega+i\omega_{\sigma})\left[(\omega^2-\omega^2_\pm)+i\omega\omega_\nu\right],
\end{equation}
\begin{equation}
\Gamma_\pm=\omega^2_\pm-3\omega^3+\omega_{\sigma}\omega_\nu-2i\omega(\omega_\nu+\omega_{\sigma}). 
\end{equation}
\end{subequations} 
These functions have transparent physical meaning. Zeroes of $\Pi_\pm$ correspond to the dispersion relations of the collective modes in the system. The first one corresponds to a spin diffusion $\omega=-i\omega_{\sigma}$, with $\omega_{\sigma}=D_\sigma q^2$. The second are the plasmon resonances $\omega=\omega_\pm$, with 
\begin{equation}\label{eq:plasmons}
\omega^2_\pm=\omega^2_q(1\pm e^{-qd}),\quad \omega_q=\sqrt{\frac{2\pi ne^2q}{\epsilon m}},
\end{equation}
whose attenuation is determined by viscous spreading of charge $\Im\omega=\omega_{\nu}/2$, here $\omega_\nu=\nu q^2$ with $\nu=(\eta+\zeta)/mn$ being the kinematic viscosity of the fluid. For completeness, we mention that in the expression for $\Pi_\pm$ we neglected an extra term that scales as $\propto q^4$, which is insignificant for fluctuations with long wave length. In all the analysis above, we also assumed that the systems is Galilean invariant. In systems with broken Galilean invariance plasmons can decay with the Maxwell mechanism of charge relaxation \cite{Zverevich:LTP23}. This aspect of the problem will not change the essence of our results for drag magnetoresistance. It only determines the enhancement factor of drag effect from plasmon resonances \cite{Flensberg:PRB94,Yu-Kuang:PRB95}, which is weakly sensitive to the mechanism of plasmon decay in the hydrodynamic limit.    

To close the system of required relations for the obtained solution, we need expressions for thermal averages, which follow directly from Eqs. \eqref{eq:Langevin-spin} and \eqref{eq:Langevin-visc}, namely, 
\begin{subequations}\label{eq:Averages}
\begin{equation}
\langle(\bm{q}\cdot\delta\bm{\varsigma}_\pm)(\bm{q}\cdot\delta\bm{\varsigma}_\pm)\rangle=4T\sigma_s q^2,
\end{equation}
\begin{equation}
\langle(\bm{q}\cdot(\bm{q}\cdot\delta\hat{\bm{\Xi}}_\pm))(\bm{q}\cdot(\bm{q}\cdot\delta\hat{\bm{\Xi}}_\pm))\rangle=4T(\eta+\zeta)q^4.
\end{equation}
\end{subequations}
Owing to the fact that random Langevin fluxes of spin currents are uncorrelated with the corresponding fluxes of viscous stresses, dynamical density structure factor $\mathcal{D}(\bm{q},\omega)$ is an additive function of these sources of density fluctuations. The same conclusion holds for the drag resistance. Therefore, these contributions can be investigated separately. 

%#################################################################################################################
%#################################################################################################################
%#################################################################################################################
%#################################################################################################################
%#################################################################################################################

\section{Spin drag magnetoresistance}\label{sec:Drag} 

We present drag resistance as a sum of two contributions 
\begin{equation}\label{eq:rho-sum}
\rho_{\text{D}}=\rho_\nu+\rho_\sigma.
\end{equation}
The first one ($\rho_\nu$) arises from density fluctuations driven by random viscous stresses. The second ($\rho_\sigma$) arises from the physics associated with the spin polarization and electron density fluctuations driven by spin current fluctuations. We focus on the spin contribution. 

%%%%%%%%%%%%%%%%%%%%%%%%%%%%%%
%%%%%%%%%%%%%%%%%%%%%%%%%%%%%%
\begin{figure}[t!]
\includegraphics[width=0.9\linewidth]{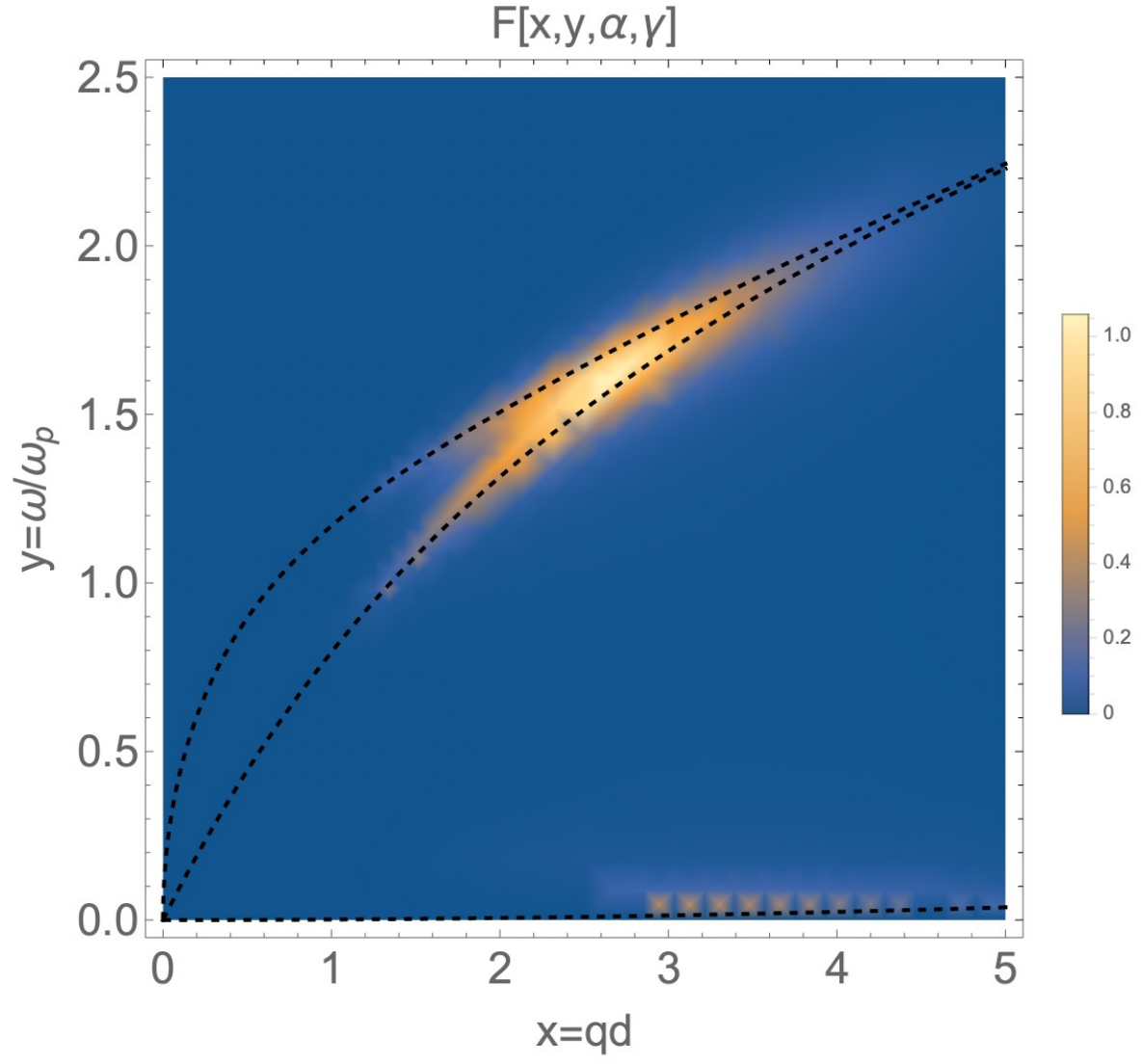}
\caption{Dispersion laws $\omega=\omega_\pm(q)$ for the optical and acoustic plasmons Eq. \eqref{eq:plasmons}, shown by dashed lines,
superimposed on top of the color plot that defines magnitude of the integrand in Eq. \eqref{eq:rho-D-spin-int}, which was normalized to itself at the maximal value. 
The lower dashed line defines broadening of the spectral weight because of spin diffusion. This plot was generated for values $\alpha=1.5\times 10^{-3}$ and $\gamma=2.5\times10^{-2}$.}
\label{fig:plasmons}
\end{figure}
%%%%%%%%%%%%%%%%%%%%%%%%%%%%%%
%%%%%%%%%%%%%%%%%%%%%%%%%%%%%%

To the linear order in $\bm{v}$, the dynamic density response function averaged over the spin fluctuations $\mathcal{D}_\sigma=\langle\delta n_1\delta n_2\rangle_\varsigma$ contains terms of the same $\langle\delta n^{(0)}_\pm\delta n^{(1)}_\pm\rangle_\varsigma$ and mixed symmetry $\langle\delta n^{(0)}_\pm\delta n^{(1)}_\mp\rangle_\varsigma$ with respect to optical and acoustic plasmon resonances. 
The products with interchanged superscript $(0)\leftrightarrow(1)$ are also possible. These products contribute with the opposite sign as $\delta n^{(1)}_\pm$ is odd in $\bm{q}$. A careful inspection of these terms reveals that they have different parity in frequency dependence. The former are frequency odd, while the latter are frequency even. Therefore, only averages of the products of mixed parity density modes contribute to the drag force in Eq. \eqref{eq:F-D}. Keeping all the relevant terms and after some straightforward algebra, we arrive at the intermediate expression  
\begin{equation}\label{eq:D-spin}
\mathcal{D}_\sigma=\frac{i(\bm{q}\cdot\bm{v})}{4}\Upsilon_q\frac{\Im(\Gamma_+\Pi^*_-)-\Im(\Gamma_-\Pi^*_+)}{|\Pi_+|^2|\Pi_-|^2},
\end{equation} 
where we introduced a short-hand notation $\Upsilon_q=(4T\sigma_s q^2)(q^2/m)^2(\partial_\sigma P)^2$. To derive this result, we used complex-conjugation properties of the polarization and coupling functions: $\Pi_\pm(-\omega)=-\Pi^*_\pm(\omega)$ and $\Gamma_\pm(-\omega)=\Gamma^*_\pm(\omega)$, and applied equilibrium averages
\begin{equation}
\left\langle \delta n^{(0)}_\pm(\bm{q},\omega)\delta n^{(0)}_\pm(-\bm{q},-\omega)\right\rangle_\varsigma=\frac{\Upsilon_q}{|\Pi_\pm|^2}
\end{equation}
that follow from Eqs. \eqref{eq:delta-n-sol} and \eqref{eq:Averages}. The numerator of Eq. \eqref{eq:D-spin} can be easily calculated with the help of Eq. \eqref{eq:Pi-Gamma} and equals to $\Im(\Gamma_+\Pi^*_-)-\Im(\Gamma_-\Pi^*_+)=3\omega^2\omega_\nu(\omega^2_+-\omega^2_-)$. Inserting now Eq. \eqref{eq:D-spin} into Eq. \eqref{eq:F-D}, and using Eq. \eqref{eq:rho-D-def} we find for the spin drag resistance 
\begin{equation}\label{eq:rho-D-spin-int}
\rho_\sigma=\frac{3}{8e^2n^2}\int\left(\frac{2\pi e^2}{\epsilon q}\right)e^{-qd}\frac{q^2\Upsilon_q\omega^2\omega_\nu(\omega^2_+-\omega^2_-)}{|\Pi_+|^2|\Pi_-|^2}d\Gamma_{q\omega}.
\end{equation}
The frequency integral here can be calculated using residues in the complex plane of $\omega$, but this approach results in a rather cumbersome outcome. Instead, we found an approximation that, as can be shown numerically, is fairly accurate and allows us to make further analytical progress. This approximation is based on the following observation. On Fig. \ref{fig:plasmons} we plot the integrand of Eq. \eqref{eq:rho-D-spin-int}, which comprises of the product of the dynamical structure factor, Coulomb potential, and the phase space factor. On the plot we introduced dimensionless variables $x=qd$ and $y=\omega/\omega_{\text{p}}$, where $\omega_{\text{p}}=\sqrt{2\pi ne^2/\epsilon md}$ is the characteristic plasmon frequency. This plot depends on two dimensionless parameters $\alpha=D_\sigma/(\omega_{\text{p}}d^2)$ and $\gamma=\nu/(\omega_{\text{p}}d^2)$. In this units, plasmon dispersions are given by $\sqrt{x(1\pm e^{-x})}$, which are shown by the dashed lines on the plot. The lower dashed line, at smallest frequencies, corresponds to the spin diffusion branch of the spectrum. It is clear that the integrand has maximal spectral weight at the plasmon poles, which is highlighted by the color-bar scale.   

Based on these considerations, we conclude that the most important regions of frequency integration in Eq. \eqref{eq:rho-D-spin-int} are those near plasmon resonances $\omega_\pm$. We have verified that the frequency region $\omega\sim\omega_\sigma$ gives smaller contribution. For typical wave numbers, $q\sim1/d$, plasmon energies are much bigger than the energy scale characteristic of spin diffusion $\omega_\pm\gg\omega_\sigma$. Therefore, the Lorentzian, $(\omega^2+\omega^2_\sigma)^{-1}$, in each of the factor $|\Pi_\pm|^{-2}$, can be approximately replaced by $\omega^{-2}_\pm$. The remaining part of the frequency integral can be calculated exactly in a compact form   
\begin{align}
\int^{+\infty}_{-\infty}\frac{\omega^2d\omega}{[(\omega^2-\omega^2_+)^2+\omega^2\omega^2_\nu][(\omega^2-\omega^2_-)^2+\omega^2\omega^2_\nu]}\nonumber \\ 
=\frac{2\pi/\omega_\nu}{(\omega^2_+-\omega^2_-)^2+2\omega^2_\nu(\omega^2_++\omega^2_-)}.
\end{align}
This simplifies $\rho_\sigma$ to the form 
\begin{align}
\rho_\sigma=\frac{3}{8e^2n^2}\int\frac{d^2q}{(2\pi)^2}
\left(\frac{2\pi e^2}{\epsilon q}\right)e^{-qd}\left(\frac{\omega^2_+-\omega^2_-}{\omega^2_+\omega^2_-}\right)\nonumber\\ 
\times \frac{q^2\Upsilon_q}{(\omega^2_+-\omega^2_-)^2+2\omega^2_\nu(\omega^2_++\omega^2_-)}.
\end{align}
Using the explicit forms of plasmon dispersions from Eq. \eqref{eq:plasmons}, the remaining momentum integral can be reduced to a dimensionless form in terms of new integration variable $x=qd$. We find as a final result (omitting overall numerical factor for brevity):
\begin{equation}\label{eq:rho-D-spin}
\rho_\sigma\simeq\frac{\sigma_s}{e^2}\left(\frac{\epsilon v_{\text{F}}}{e^2}\right)^2\frac{f(\gamma)}{(nd^2)^3}
\frac{T}{E_{\text{F}}}\left(\frac{H}{E_{\text{F}}}\right)^2,
\end{equation}
where we introduced dimensionless function  
\begin{equation}\label{eq:f}
f(\gamma)=\int^{\infty}_{0}\frac{x^5e^{-x}dx}{\sinh(x)(e^{-2x}+\gamma^2 x^3)},
\end{equation}
and the dimensionless parameter
\begin{equation}
\gamma=\frac{r_s}{\sqrt{2}}\frac{\eta+\zeta}{n}\left(\frac{a_{\text{B}}}{d}\right)^{\frac{3}{2}}.
\end{equation}
In order to extract the magnetic field dependence, $\propto H^2$, which is implicit in $\partial_\sigma P$, we used general thermodynamic principles. The equation of state is determined by the grand thermodynamic potential $\Omega=-PV$ \cite{LL-V5}. 
Provided the linear relationship between the magnetic polarization and the field, $\sigma=\chi H$, and using the thermodynamic theorem of small corrections, one finds $\Omega_H=\Omega_0-\chi H^2/2$, where $\Omega_0$ is the potential of interacting systems without the field. From this argument, it is evident that $\partial_\sigma P\propto H$. 

The field-independent part of drag resistance, can be calculated in the same fashion. Indeed, using thermal averages 
\begin{equation}
\left\langle \delta n^{(0)}_\pm(\bm{q},\omega)\delta n^{(0)}_\pm(-\bm{q},-\omega)\right\rangle_\Xi=\frac{(\omega^2+\omega^2_\sigma)\Lambda_q}{|\Pi_\pm|^2}
\end{equation}
where $\Lambda_q=4T(\eta+\zeta)q^4/m^2$. This leads then to a corresponding part of the density structure factor computed to the linear order in $\bm{v}$,
\begin{equation}
\mathcal{D}_\nu=\frac{i(\bm{v}\cdot\bm{q})(nTq^2/m)(\omega^2_+-\omega^2_-)\omega^2_\nu}{[(\omega^2-\omega^2_+)^2+
\omega^2\omega^2_\nu][(\omega^2-\omega^2_-)^2+\omega^2\omega^2_\nu]}.
\end{equation}
Using Eqs. \eqref{eq:F-D} and \eqref{eq:rho-D-def}, performing integrations one finds the corresponding drag resistance 
\begin{equation}\label{eq:rho-D-visc}
\rho_\nu\simeq\left(\frac{\epsilon v_{\text{F}}}{e^4}\right)\frac{\eta+\zeta}{n}\frac{h(\gamma)}{(k_{\text{F}}d)^5}\frac{T}{E_{\text{F}}}, 
\end{equation} 
where the dimensionless function is given by 
\begin{equation}\label{eq:h}
h(\gamma)=\int^{\infty}_{0}\frac{x^4e^{-x}(1+\gamma^2x^3)dx}{\sinh(x)(e^{-2x}+\gamma^2 x^3)}.
\end{equation}
Drag resistance defined by Eqs. \eqref{eq:rho-D-spin} and \eqref{eq:rho-D-visc} are the main results of this paper.   

%#################################################################################################################
%#################################################################################################################
%#################################################################################################################
%#################################################################################################################
%#################################################################################################################

\section{Summary and discussion}\label{sec:Summary}

A detailed analysis of Eqs. \eqref{eq:rho-D-spin} and \eqref{eq:rho-D-visc} for the drag resistance depends essentially on specific assumptions about the density and temperature dependence of spin conductivity and electron liquid viscosity. These quantities have been calculated microscopically only in the limit of a weakly interacting Fermi gas \cite{SykesBrooker:1970}. In the context of the introduction and discussion of the observed features of drag in 2DES with large $r_s$, we can provide several comparative comments.

(i) The measured spin diffusion constant via spin Coulomb drag in 2DES \cite{Weber:2005} reveals that $D_\sigma$ is almost temperature independent in a broad range $T>E_{\text{F}}/10$. With this input, Eq. \eqref{eq:rho-D-spin} predicts that drag magnetoresistance is linear in temperature, $\rho_\sigma\propto T$. This temperature behavior is qualitatively consistent with the observations \cite{Pillarisetty:PRL03, Pillarisetty:PRL05}.

\begin{figure}[t!]
\includegraphics[width=0.9\linewidth]{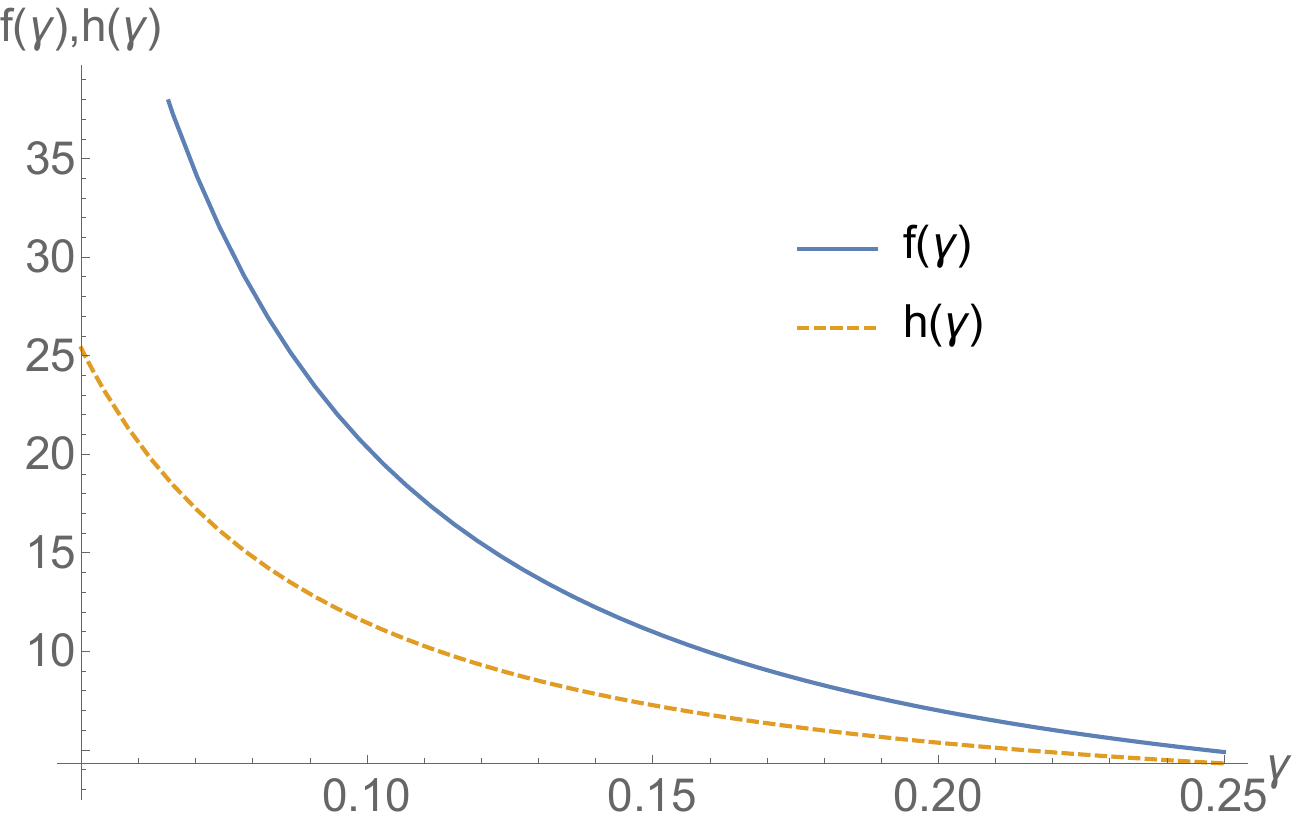}
\caption{Dependence of the dimensionless functions from Eqs. \eqref{eq:f} (solid line) and \eqref{eq:h} (dashed line) on the parameter $\gamma$ that determines broadening of the plasmon resonances and thus enhancement of the drag resistance in Eqs. \eqref{eq:rho-D-spin} and \eqref{eq:rho-D-visc} respectively. The shown plot range is $\gamma\in[0.05,0.25]$.}
\label{fig:f-h}
\end{figure}

(ii) The plasmon enhancement factor described by $f(\gamma)$ is also temperature dependent, but only logarithmically. Indeed, it can be readily verified that for $T<E_{\text{F}}$ the parameter $\gamma<1$. The asymptote of Eq. \eqref{eq:f} in the limit $\gamma\ll1$ is $f\sim 2\ln^6[\gamma\ln(1/\gamma)]$. For a Fermi liquid $\eta\sim n(E_{\text{F}}/T)^2$, therefore $f$ behaves logarithmically with $T$. Note, however, that numerically $f\sim 25$ for $\gamma\sim 0.1$ so that plasmon enhancement is significant and strongest at matched density between the layers. The corresponding enhancement for the field independent part, as described by $h(\gamma)$ function in Eq.\eqref{eq:h}, is slightly weaker, see Fig. \ref{fig:f-h} for comparison.   

(iii) Equation \eqref{eq:rho-D-spin} also predicts quadratic field dependence, $\rho_\sigma\propto H^2$, which we expect to saturate at higher fields since spin polarization saturates. Additional complications may arise from the $H$ dependence of the kinetic coefficients $\eta(H), \zeta(H)$. 

(iv) As a function of density, the relevant prefactor in Eq. \eqref{eq:rho-D-spin} scales as $\rho_\sigma\propto \sigma_s(n)/n^5$. Provided that spin conductivity does not change significantly in the measured range of densities, the power law scaling of drag resistance with $n$ is therefore also in good agreement with the observed dependence.   

(v) The comparison of Eq. \eqref{eq:rho-D-visc} to the FL formula for drag resistivity Eq. \eqref{eq:rho-D-FL} reveals that intralayer collisions strongly enhance drag effect. This is qualitatively consistent with the fact that drag resistance is bilayers with large $r_s$ is found to be two to three orders of magnitude larger than expected on the basis of FL theory \cite{Pillarisetty:PRL02}.  

(vi) The quantum Monte Carlo results for the spin susceptibility $\chi$ reported \cite{Bachelet:PRL02} have been widely compared with experiment e.g. \cite{Zhu:PRL03}, and appear quite reliable and accurate. These studies show that spin susceptibility of a 2DES, measured in the units of Pauli susceptibility, $\chi/\chi_{\text{P}}$, grows by an order of magnitude when $r_s$ is increased. As a consequence, based on the Einstein relation $\sigma_s=\chi D_\sigma$, Eq. \eqref{eq:rho-D-spin} predicts strongly enhanced drag magnetoresistance.    

(vii) We note that our results can be extrapolated to the nondegenerate limit $T>E_{\text{F}}$ that includes both semiquantum and strongly correlated classical plasma. For example, for a classical limit $\eta\sim\zeta\sim mv_TT/e^2$, where $v_T\sim\sqrt{T/m}$ is the thermal velocity, so that $\rho_\nu\propto T^{5/2}$.  

%%%%%%%%%%%%%%%%%%%%%%%%%%%%%%
%%%%%%%%%%%%%%%%%%%%%%%%%%%%%%
%%%%%%%%%%%%%%%%%%%%%%%%%%%%%%
%%%%%%%%%%%%%%%%%%%%%%%%%%%%%%

In light of the theoretical insights, it is instructive to discuss the reported puzzling similarities between drag resistivity and intralayer magnetoresistance \cite{Pillarisetty:PRL03, Pillarisetty:PRL05}. The hydrodynamic mechanism of spin magnetoresistance was proposed in Ref. \cite{AKS}. It was shown that in-plane magnetoresistance is primarily determined by spin diffusion, whereas zero-field resistance is dominated by the viscous term, which is similar to our result for drag. This similarity is not accidental and has clear physical interpretation. Indeed, in the model of Ref. \cite{AKS}, electron flow occurs in the presence of a long-range disorder potential with a correlation radius exceeding the electron mean free path. In this context, intralayer magnetoresistivity can be understood in terms of the drag force between the electron liquid and the disorder potential. Specifically, the disorder potential creates fluctuations in the particle and spin density within the electron liquid. To second order in the disorder potential, the subsequent scattering of these density fluctuations from the disorder potential produces a net resistive force. In the case of Coulomb drag, both the scattering potential and the fluctuations in electron density are produced by thermal fluctuations, whose variance depends on the temperature, proportional to $T$, through the strength of Langevin fluxes. This aspect of the problem accounts for the difference in the temperature dependence of drag and intralayer magnetoresistivity. On the other hand, the propagation of fluctuations in the fluid, in either case, is described by the same linearized hydrodynamic equations and occurs in the form of stress-driven viscous modes and spin fluctuation-driven diffusive modes. This results in the similarity between the corresponding transport coefficients $\rho$ and $\rho_{\text{D}}$.

The result for drag magnetoresistance in Eq. \eqref{eq:rho-sum} can be generalized to systems with broken Galilean invariance. The key modification required is the inclusion of the electrical current because of finite intrinsic conductivity $\sigma_c$ (i.e., charge conductivity relative to the fluid, often referred to as microscopic ``incoherent" conductivity). Consequently, Langevin fluxes must be augmented to account for fluctuations arising from intrinsic currents in the fluid. These fluctuations introduce an additional mechanism for density variation via current continuity, thereby contributing to the dynamical structure factor of the fluid and the interlayer drag force.
While the drag effect is still enhanced by plasmons, their lifetime is now governed by the Maxwell time of charge relaxation rather than viscous diffusion. For long-wavelength density fluctuations, the corresponding attenuation coefficient is given by $\Im\omega_q=2\pi\sigma_cq/\epsilon$. This change in physics affects the broadening of plasmon resonances and the enhancement factor in drag resistance. The analysis indicates that Eq. \eqref{eq:rho-sum} acquires an additional contribution of the form
\begin{equation}
\rho_c=\frac{\sigma_c}{4\pi^2e^4}\frac{T}{E_{\text{F}}}\frac{1}{(nd^2)^2}g(\beta). 
\end{equation}
The dimensionless function $g(\beta)$ captures the plasmon enhancement and exhibits a logarithmic dependence on the dimensionless parameter
\begin{equation}
\beta\simeq\frac{\sigma_c}{e^2}\sqrt{\frac{e^2}{\epsilon v_{\text{F}}}}\frac{1}{\sqrt{k_{\text{F}}d}}.
\end{equation}
The main parametric dependence in Eqs. \eqref{eq:rho-D-spin} and \eqref{eq:rho-D-visc} remains the same; however, the dimensionless functions $f(\gamma)\to f(\beta)$ and $h(\gamma)\to h(\beta)$ are described by different analytic expressions in terms of $\beta$ as compared to Eqs. \eqref{eq:f} and \eqref{eq:h} respectively. Therefore, modulo logarithmic factors, the general conclusion regarding the field and temperature dependence of the drag magnetoresistance remains the same in this case as well.

\section*{Acknowledgments}

We are grateful to Anton Andreev, Steven Kivelson, and Boris Spivak for insightful discussions. This research project was financially supported by the National Science Foundation (NSF) Grant No. DMR-2203411 and H. I. Romnes Faculty Fellowship provided by the University of Wisconsin-Madison Office of the Vice Chancellor for Research and Graduate Education with funding from the Wisconsin Alumni Research Foundation. I.E. was supported by the University of Wisconsin -- Madison. This paper was finalized at the Aspen Center for Physics, during the program ``Quantum Matter Through the Lens of Moir\'e Materials", which is supported by the NSF Grant No. PHY-2210452. 

\bibliography{Spin-Drag-Biblio}

\end{document}